\documentclass[fleqn,usenatbib]{mnras}

\usepackage{newtxtext,newtxmath}
\usepackage[T1]{fontenc}

\DeclareRobustCommand{\VAN}[3]{#2}
\let\VANthebibliography\thebibliography
\def\thebibliography{\DeclareRobustCommand{\VAN}[3]{##3}\VANthebibliography}

\usepackage{graphicx}	% Including figure files
\usepackage{amsmath}	% Advanced maths commands
\usepackage{scalerel}
\usepackage{tikz}
\usetikzlibrary{svg.path}
\definecolor{orcidlogocol}{HTML}{A6CE39}
\tikzset{
  orcidlogo/.pic={
    \fill[orcidlogocol] svg{M256,128c0,70.7-57.3,128-128,128C57.3,256,0,198.7,0,128C0,57.3,57.3,0,128,0C198.7,0,256,57.3,256,128z};
    \fill[white] svg{M86.3,186.2H70.9V79.1h15.4v48.4V186.2z}
                 svg{M108.9,79.1h41.6c39.6,0,57,28.3,57,53.6c0,27.5-21.5,53.6-56.8,53.6h-41.8V79.1z M124.3,172.4h24.5c34.9,0,42.9-26.5,42.9-39.7c0-21.5-13.7-39.7-43.7-39.7h-23.7V172.4z}
                 svg{M88.7,56.8c0,5.5-4.5,10.1-10.1,10.1c-5.6,0-10.1-4.6-10.1-10.1c0-5.6,4.5-10.1,10.1-10.1C84.2,46.7,88.7,51.3,88.7,56.8z};
  }
}

\newcommand\orcidicon[1]{\href{https://orcid.org/#1}{\mbox{\scalerel*{
\begin{tikzpicture}[xscale=1,yscale=-1, transform shape]
\pic{orcidlogo};
\end{tikzpicture}
}{|}}}}

\usepackage{threeparttable}

\newcommand{\solarM}{\,\mathrm{M}_\odot}

\newcommand{\E}[1]{\times10^{#1}}

\newcommand{\ergs}{\,\mathrm{erg}\,\mathrm{s}^{-1}}

\newcommand{\LX}{L_\mathrm{X}}
\newcommand{\LEdd}{L_\mathrm{Edd}}

\newcommand{\xmm}{\emph{XMM-Newton}}

\newcommand{\chandra}{\emph{Chandra}}

\newcommand{\einstein}{\emph{Einstein Observatory}}

\title[Magnetically-driven QPOs in PULXs]{%Pulsating ULX time variability as probe for magnetically-driven precession \\
Wobbling around the clock: magnetically-driven quasi-periodic oscillations in pulsating ultraluminous X-ray sources
}

\author[M. Veresvarska et al.]{M. Veresvarska,$^{\orcidicon{0000-0002-0146-3096}}$$^{1}$\thanks{E-mail: martina.veresvarska@durham.ac.uk}\thanks{These authors contributed equally.}
M. Imbrogno,$^{\orcidicon{0000-0001-8688-9784}}$$^{2,3,4}$\thanks{E-mail: matteo.imbrogno@inaf.it}\footnotemark[2]
R. Amato,$^{\orcidicon{0000-0003-0593-4681}}$$^{3}$
G.~L. Israel,$^{\orcidicon{0000-0001-5480-6438}}$$^{3}$
S. Scaringi,$^{\orcidicon{0000-0001-5387-7189}}$$^{1,5}$
\newauthor
P. Casella,$^{\orcidicon{0000-0002-0752-3301}}$$^{3}$
D. de Martino,$^{\orcidicon{0000-0002-5069-4202}}$$^{5}$,
F. F{\"urst},$^{\orcidicon{0000-0003-0388-0560}}$$^{6}$,
A. G\'urpide,$^{\orcidicon{0000-0002-2256-2704}}$$^{7}$,
C. Knigge$^{\orcidicon{0000-0002-1116-2553}}$$^{7}$,
M. J. Middleton,$^{7}$
\\
$^{1}$Department of Physics, Centre for Extragalactic Astronomy, Durham University, South Road, Durham DH1 3LE, UK\\
$^{2}$Dipartimento di Fisica, Università degli Studi di Roma “Tor Vergata”, via della Ricerca Scientifica 1, I-00133 Rome, Italy\\
$^{3}$INAF -- Osservatorio Astronomico di Roma, via Frascati 33, I-00078 Monte Porzio Catone (RM), Italy\\
$^{4}$Dipartimento di Fisica, Università degli Studi di Roma “La Sapienza”, piazzale Aldo Moro 5, I-00185 Roma, Italy\\
$^{5}$INAF -- Osservatorio Astronomico di Capodimonte, Salita Moiariello 16, I-80131 Naples, Italy\\
$^{6}$European Space Astronomy Centre (ESAC), ESA, Camino Bajo del Castillo s/n, Villanueva de la Ca\~{n}ada, E-28692 Madrid, Spain\\
$^{7}$School of Physics and Astronomy, University of Southampton, Highfield, Southampton SO17 1BJ, UK
}

\date{Accepted XXX. Received YYY; in original form ZZZ}

\pubyear{\the\year{}}

\begin{document}
\label{firstpage}
\pagerange{\pageref{firstpage}--\pageref{lastpage}}
\maketitle

% Abstract of the paper
\begin{abstract}
Ultraluminous X-ray sources (ULXs) are X-ray binary systems containing an accreting neutron star (NS) or black hole emitting at luminosities above the Eddington limit of a $10\solarM$ black hole. Approximately 1900 (either confirmed or candidate) ULXs have been identified to date. Three systems have been confirmed to exhibit coherent signals consistent with NS spin frequencies and quasi-periodic oscillations (QPOs) in the mHz range. Several interpretations for generating such QPOs have been proposed, including general relativistic frame-dragging effects. In this work, we test if an alternative model in which magnetically-driven precession of the inner accretion flow can self-consistently reproduce the observed NS spin and QPO frequencies for reasonable values for accretion rates and NS magnetic field strengths. For a range of  parameters, we recover family of solutions with accretion rates $\approx10^{-7}$--$10^{-5}$\,M$_{\odot}$\,yr$^{-1}$ and surface magnetic fields $\gtrsim10^{12}$\,G, in agreement with previous estimates. If validated, this interpretation could reconcile several observed properties of pulsating ULXs, including QPO frequencies and the observed high luminosities of these systems, in a self-consistent framework without requiring general relativistic effects and/or strong beaming due to specific viewing angles. Although the predictive power of the model is currently limited by parameter degeneracies and uncertainties, searching for and discovering more pulsating ULX systems will allow to further test or refute the proposed model.

\end{abstract}

% Select between one and six entries from the list of approved keywords.
% Don't make up new ones.

%%%%%%%%%%%%%%%%%%%%%%%%%%%%%%%%%%%%%%%%%%%%%%%%%%

%%%%%%%%%%%%%%%%% BODY OF PAPER %%%%%%%%%%%%%%%%%%

\section{Introduction}
\label{sec:Introduction}

Ultraluminous X-ray sources (ULXs) are a class of X-ray binaries with luminosities $\LX\gtrsim10^{39}\ergs$, exceeding the Eddington limit ($\LEdd\simeq1.3\E{38}M/M_\odot\ergs$, with $M$ being the mass of the accretor) of a $\sim$10\,M$_\odot$ black hole \citep[BH; see][for recent reviews]{Kaaret2017,Fabrika2021,King2023,Pinto2023a}. First detected by the \einstein\ in the off-nuclear regions of nearby galaxies \citep{Fabbiano1989}, they have been proposed to be possible intermediate-mass black hole (IMBH) candidates with $M_\mathrm{BH}\simeq10^2-10^6\solarM$, accreting at sub-Eddington rates \citep[see e.g.][]{Colbert1999}. The presence of mHz quasi-periodic oscillations (QPOs) in some ULXs \citep[see e.g.][]{Strohmayer2007,Strohmayer2009,Pasham2015,Atapin2019}, interpreted as the low-frequency counterparts of QPOs in Galactic BH binaries \citep{vanderKlis1989}, seemingly supported this hypothesis. Indeed, using the mass-frequency scaling derived from Galactic BH binaries \citep[e.g.][]{Aschenbach2004,Remillard2006,Smith2018}, the masses inferred from the QPO frequencies are found to be in the IMBH range (see e.g. \citealp{Casella2008} and Figure 4 of \citealp{Smith2018}). However, some differences cast doubt on such simple analogies. For example, in the case of Galactic BH binaries, the QPO frequency $\nu_\mathrm{QPO}$ tightly correlates with a low frequency break $\nu_\mathrm{b}$, with $\nu_\mathrm{QPO}\approx10\nu_\mathrm{b}$ \citep{Wijnands1999}. Such a correlation is not observed in ULXs with QPOs \citep[see e.g.][]{Middleton2011}.

The detection in a few ULXs of coherent pulsations with $P\sim0.1-10$\,s -- known as pulsating ULXs (PULXs) -- has demonstrated that at least some of these sources are powered instead by neutron stars (NSs) accreting at super-Eddington rates \citep{King2001,Poutanen2007,Zampieri2009}. A few of these PULXs also show mHz QPOs at super-Eddington luminosities, providing additional evidence that care is needed when using the QPO frequency in a (P)ULX to estimate the mass of the accretor. A total of 12 (either confirmed or candidate) PULXs have been discovered \citep[see Table~2 of][]{King2023}, with luminosities up to $\sim10^{41}\ergs$ \citep{Israel2017a}. How these sources can reach such high luminosities is still unclear, with some arguing that the inferred luminosities (derived under the isotropic emission approximation) could be overestimated due to  %the real luminosities could be much lower and 
%artificially induced by 
geometrical beaming \citep[see e.g.][]{King2017,Lasota2023}. 

In this paper, we focus on those PULXs that have shown mHz QPOs at (observed) luminosities $\gtrsim10^{39}\ergs$: M82\,X-2 \citep{Feng2010}, M51\,ULX-7 \citep{Imbrogno2024} and NGC\,7793 P13 (Imbrogno et al., in prep.). M82\,X-2 was the first source identified as a PULX \citep{Bachetti2014}. With a luminosity $\LX\simeq10^{39}-10^{40}\ergs$ and an orbital period $P_\mathrm{orb}\simeq2.5$\,d, the NS powering M82\,X-2 is found near spin equilibrium, with a spin period $P\simeq1.37$\,s. The source alternates between strong spin-up and spin-down phases over the years, despite less drastic changes in luminosity, and shows a spin evolution inconsistent with that expected from a slow rotator, as shown by \cite{Bachetti2020}. M82\,X-2 is also a good example of how the estimate of the mass of the accretor in a ULX from the QPO frequency can be misleading. The nature of the accretor was unknown when \cite{Feng2010} detected a QPO at $\nu\simeq3$\,mHz in a few \chandra\ observations, leading the authors to conclude that the source was a good IMBH candidate, with $M\simeq12000-43000\solarM$, $\sim$ four orders of magnitude larger than the real mass ($M\simeq1.4\solarM$). 

M51\,ULX-7, whose spin pulsations at $P\simeq2.8$\,s were first detected by \cite{RodriguezCastillo2020}, is the PULX with the shortest known orbital period \citep[$P_\mathrm{orb}\simeq2$\,d;][]{Hu2021,Vasilopoulos2021}. It is also characterised by a super-orbital modulation, initially detected with a period $P_\mathrm{so}\simeq38$\,d, but later found to be gradually evolving towards a longer period $P_\mathrm{so}\simeq45$\,d \citep{Brightman2020,Vasilopoulos2020a,Brightman2022}. Recently, \cite{Imbrogno2024} detected a flaring-like feature in the light curve of three \xmm\ observations of the source. A Fourier analysis of these data revealed the presence of a QPO at a frequency $\nu\simeq0.5$\,mHz. The same feature was detected also in \chandra\ archival observations. 
%The variability is linked with a QPO at $\nu\simeq0.5$\,mHz, detected also in \chandra\ archival observations. 
In both sources, the QPO is always present at the same frequency 
%, suggesting that it is linked with a specific region of the system
\citep{Imbrogno2024}. 

Lastly, NGC\,7793 P13 is the fastest known PULX, with a spin period $P\simeq0.4$\,s \citep{Furst2016,Israel2017}. It is the only PULX with an identified optical counterpart \citep{Motch2011}. A peculiarity of this system is that the orbital periods estimated through optical ($P_\mathrm{opt}\simeq64$\,d) and X-ray ($P_\mathrm{X}\simeq65$\,d) are not compatible, an inconsistency which remains without a clear explanation \citep{Furst2018,Furst2021}. Recently, a QPO at $\nu\simeq10$\,mHz was detected in a few \xmm\ observations of NGC\,7793 P13 (Imbrogno et al., in prep.). As for M51\,ULX-7, the QPO is always found at a specific frequency and only when the system emits at super-Eddington luminosities.  

Various models have been proposed to explain the presence of mHz QPOs in (P)ULXs. \cite{Middleton2019} proposed that mHz QPOs arise from a precessing inner flow of the disc. The general relativistic frame-dragging torque (inducing Lense-Thirring precession) is then communicated to the launched winds, whose precession is expected to cause the much longer super-orbital period seen in many (P)ULXs \citep[see e.g.][]{Kong2016,Walton2016a,Furst2018,Vasilopoulos2020a,Brightman2020,Brightman2022}. \cite{Atapin2019}, instead, proposed that the propagating fluctuations mechanism from \cite{Lyubarskii1997} can explain both the QPO and the flat-topped noise observed in a sample of ULXs. \cite{Majumder2023} followed \cite{Das2021} to link the QPO frequency to the infall time towards the inner radius of the disc, and hence proposing to use the QPO frequency to infer the masses of the central object, assumed to be a BH. A similar mechanism for generating quasi-periodic variability, but for NSs, is also explored in \citet{Mushtukov2024MNRAS.530..730M}.

Contrary to BHs, the presence of a magnetosphere and its dynamical importance to the accretion flow in PULXs can be substantial and is expected to exert additional torques on the inner flow. The effect of the magnetosphere through what we refer to here as the magnetically-driven precession model (MDP model hereafter) has been explored for both magnetised NSs and T Tauri stars by \citet{Lai1999} and more recently for accreting white dwarfs (WDs) by \citet{Veresvarska2024}. In this paper, we explore the validity of applying the MDP model, presented in Sect.~\ref{sec:Model}, to the mHz QPOs recently detected in a few PULXs. In particular, we want to verify if the MPD model can simultaneously match the spin and QPO periods observed in these PULXs for reasonable values of the accretion rate and the magnetic field. With this goal in mind, we apply the model to the PULXs showing QPOs and present the results in Sect.~\ref{sec:Results}. Finally, we discuss the implications of our results, together with the strengths and limitations of the model in Sect.~\ref{sec:Discussion}.

\section{Magnetic precession model}
\label{sec:Model}
% An alternative model for explaining QPOs in neutron stars and T Tauri stars was introduced by \citet{Lai1999} and later extended to accreting white dwarfs by \citet{Veresvarska2024}.
The MDP model can in principle generate QPOs from magnetically driven precession, where the accretion flow surrounding a rotating magnetised star experiences a quasi-periodic wobbling motion around the star's spin axis. The interaction between the disc and the star's magnetic field induces a warping effect, causing the inner flow to deviate from its equatorial plane and to undergo precession. This precessing motion is driven by the torque that arises from the interaction between the misaligned disc's surface currents to the star's magnetic field, generated by its dipole, in the plane of the accretion disc.

The MDP model further explores the non-linear evolution of the warped disc, as detailed by \citet{Pfeiffer2004}. The application of this model to low-frequency QPOs in NSs is also examined in \citet{Shirakawa2002,Shirakawa2002a}. Within this framework the global precession of the disc can generate the observed QPOs, offering insights into the underlying mechanisms driving these oscillations in various accreting systems, including WDs and NSs.

%The QPO is attributed to the precession of the inner accretion flow around the spin axis of the central object. 
\citet{Lai1999} estimates the precession frequency of the entire inner flow by scaling the magnetic precession frequency of a specific ring at a characteristic radius, $r$, $\nu_{\mathrm{p}}(r)$, with a dimensionless constant $A$. The value of the constant $A$ depends on the disc structure. \cite{Shirakawa2002,Shirakawa2002a} estimate $A$ in the range 0.3$-$0.85 and here we fix this to a fiducial value of 0.65 (close to the midpoint of the quoted range). This leads to the expression for the QPO frequency:
\begin{equation} \nu_\mathrm{QPO} = A \nu_p(r) = \frac{ A }{ 2\pi^3 } \frac{ \mu^2 }{ r^7~\Omega(r)~\Sigma(r) } \frac{ F(\theta) }{ D(r) }, \label{eq
} \end{equation}

\noindent where $\mu$ represents the stellar magnetic dipole moment ($\mu = B R^3$, with $B$ as the surface magnetic field strength and $R$ the stellar radius), $\Omega$ is the Keplerian angular frequency, and $\Sigma$ is the surface density of the disc (here assumed to be represented by Equation 5.41 in \citep{Frank2002apa..book.....F}) as adopted for the NS application of the MDP model in \citet{Lai1999,Shirakawa2002a}.

The dimensionless function $F(\theta)$ is defined such that $F \left( \theta \right) = 2 f \cos^{2} \theta - \sin^{2} \theta$, where $\theta$ is the angle between the magnetic moment of the accretor and the angular momentum of the disc. The dimensionless number $f$ ($0 \leq f \leq 1$) determines what part and how much of the vertical magnetic field is being screened out; here, we take $f=0$ as in \citet{Pfeiffer2004,Veresvarska2024}, so that only the spin-variable vertical field is screened out (for $f = 1$ all the vertical field is screened out). $D \left( r \right)$ is a dimensionless function given by
\begin{equation}
    D \left( r \right) = \mathrm{max} \left( \sqrt{\frac{r^{2}}{r_\mathrm{in}^{2}} - 1} , \sqrt{\frac{2 H \left( r \right)}{r_\mathrm{in}}} \right),
    \label{eq:Dr}
\end{equation}

\noindent where $H \left( r \right)$ is the half-height of the disc at radius $r$ and $r_\mathrm{in}$ is the magnetospheric radius from Eq. \ref{eq:Rm}. In the first applications to NS in \citet{Lai1999,Shirakawa2002a} and the WD application in \citet{Veresvarska2024} of the MDP model the disc half-height is assumed from Equation 5.41 in \citet{Frank2002apa..book.....F}. Here we retain this height prescription. The overall sub-Eddington treatment of the model here is anchored in the assumption that $r_\mathrm{M} > r_\mathrm{sph}$, where $r_\mathrm{sph}$ is the spherisation radius, defined as the radius at which the inflow starts to be supercritical \citep{Poutanen2007} and $r_\mathrm{M}$ is the magnetospheric radius.

An alternative half height prescription which could lead to exploration of the model within the spherisation radius is detailed in \citep{Lipunova1999AstL...25..508L}. Compared to the thin disc prescription, the steeper dependence of disc height on accretion rate results in a $\frac{H}{R} \sim 1$ at $\dot{M} \sim 10^{-6} - 10^{-5} M_{\odot}yr^{-1}$ for a 1.4$M_{\odot}$ NS with 10~km radius. This prescription for the half-height of the disc in \citet{Lipunova1999AstL...25..508L} results in the inverse dependence of QPO frequency on accretion rate as opposed to the direct dependence for a thin-disc prescription. However, implementation of the super-Eddington regime would also require further changes to the model, namely in the surface density and the condition of $r_\mathrm{M} < r_\mathrm{sph}$. Exploration of this implementation is the subject of future work and beyond the scope of this paper.

Table \ref{tab:model_par} lists all the parameters adopted in this study together with the explored ranges, where some are set to fixed fiducial values. Although some of the parameter ranges can be constrained from the work of \citet{Chashkina2017,Chashkina2019} (where $\alpha \sim 0 - 0.2$ for M82 X-2), we here explore all parameter ranges given in Table \ref{tab:model_par} for completeness.

In order to reduce several parameter degeneracies, we here assume spin equilibrium such that the co-rotation ($r_\mathrm{co}$) and magnetosphere ($r_\mathrm{M}$) radii are equal. This is a reasonable assumption in the case of M82\,X-2, as already discussed in the Introduction, while it comes with some caveats for NGC\,7793 P13 and M51\,ULX-7, which are currently spinning up (see \citealp{Furst2021,Furst2025,Brightman2022} and the Discussion).
Assuming spin equilibrium, we can anchor the QPO radius to the spin and magnetic field, reducing the degeneracy between them.  Thus, $r = r_\mathrm{M} = r_\mathrm{co}$, where the co-rotation radius is defined as:
\begin{equation}
    r_{\mathrm{co}} = \left( \frac{GMP_{\mathrm{spin}}^{2}}{4 \pi^{2}} \right) ^{\frac{1}{3}}
    \label{eq:Rc},
\end{equation}

\noindent where $P_{\mathrm{spin}}$ is the spin period of the NS, $M$ its mass and $G$ the gravitational constant. Similarly, the magnetospheric radius is here defined as:
\begin{equation}
    r_{\mathrm{M}} = \eta \left( \frac{2 \pi^{2}}{\mu_{0}^{2}} \frac{\mu^{4}}{G M \dot{M}^{2}} \right) ^{\frac{1}{7}},
    \label{eq:Rm}
\end{equation}

\noindent where $\mu_{0}$ is the vacuum permeability. $\eta$ is generally assumed to be 0.5 \citep{Wang1987,Ghosh1979a,Campana2018}, however a wider range of values (Table \ref{tab:model_par}) are explored here. Implications of this assumption on the model are further discussed in Section \ref{sec:Discussion}. 

A parameter of the MDP model is the mass accretion rate $\dot{M}_\mathrm{acc}$ onto the compact object, which can differ from the mass loss rate from the donor star through the L1 point (see  Section \ref{sec:Discussion}). We can obtain a rough estimate of $\dot{M}_\mathrm{acc}$ through the gravitational energy released assuming most is emitted at X-ray wavelengths: 

\begin{equation}
    \dot{M}_\mathrm{acc}=\frac{2 b R_\mathrm{NS} L_\mathrm{obs}}{GM_\mathrm{NS}}
    \label{eq:mdot}
\end{equation}
where $b$ is the beaming factor (such that $b=1$ corresponds to no beaming), $L_\mathrm{obs}$ is the observed isotropic luminosity of the source, $M_\mathrm{NS}$ and $R_\mathrm{NS}$ are the mass and radius of the NS. Note that the $L_\mathrm{obs}$ does not distinguish between the luminosity of the different components of the system (i.e. accretion disc, winds, etc.), but represents the observed luminosity as an upper limit. Furthermore, the linear relation between accretion rate and luminosity is used, as opposed to Equation 4 in \citep{Middleton2023}, since within the framework of the assumptions made here, it is expected that $r_\mathrm{M} > r_\mathrm{sph}$. In such cases super-Eddington accretion can be sustained for high magnetic field strengths \citep{Gurpide2021}.

%For parameters whose values cannot be reasonably fixed, such as the strength of the magnetic field of the NS and the accretion rate a parameter range explored in Figure \ref{fig:QPO_model} is given.

\begin{table}
	\centering
	\caption{Model parameters for the magnetically driven precession model for QPOs in NS XRBs. Parameters with their explored ranges are given. For some cases they are fixed to a fiducial value, where only that one is given and denoted by $^{*}$.}
	\label{tab:model_par}
 %\resizebox{\columnwidth}{!}{
	\begin{tabular}{lr} % four columns, alignment for each
		\hline
		Model Parameter& Value \\
		\hline
		M$^{*}$ ($M_{\odot}$) & 1.4 \\
        R$^{*}$ (km)  & 10 \\
        B (G) & 10$^{8}$ $-$ 10$^{15}$  \\
        $\dot{M}$ ($M_{\odot}yr^{-1}$)& 10$^{-10}$ $-$ 10$^{-3}$  \\
        $\alpha$ & 10$^{-3}$ $-$ 1\\
        $\eta$  & 10$^{-2}$ $-$ 1 \\
        $\theta$ ($^{\circ}$) & 0 - 90 \\
        A$^{*}$  & 0.65 \\
		\hline
	\end{tabular}
 %}
\end{table}

\begin{table*}
 \label{tab:2}
\begin{center}
 \caption{List of observational parameters present in literature and the B field inferred form the MDP model, for the three PULXs considered in this work}
\begin{threeparttable}
%\resizebox{\textwidth}{!}{
 \begin{tabular}{lccccccc}
  \hline
  Source & References & $P_\mathrm{spin}$ & $\nu_\mathrm{QPO}$ & $-\dot{M}_{\star}$ & $\dot{M}_{\mathrm{acc}}/b$ & $b$ &   $B_{\mathrm{MDP}}$ \\
  & & (s) & (mHz) & \multicolumn{2}{c}{(10$^{-7}$\,M$_{\sun}$\,yr$^{-1}$)}
%   & (10$^{-7}$\,M$_{\sun}$\,yr$^{-1}$)& 
& &($10^{13}$\,G) \\
  \hline
%  \multicolumn{7}{c}{$\alpha$=0.25, $\kappa$=0.5} \\
%  \hline
M82 X-2 & [1,2,3] & 1.32$-$1.39 & 2.77$-$3.98 & 47(2) & 20$-$40& 1& $>0.8$  \\
&&&&&2$-$4&0.1 & $>0.3$ \\
\hline
M51 ULX-7 & [4,5,6] & 2.79$-$3.28 & 0.53$-$0.56 & -- & 5$-$12 & 1& $>2$  \\
   %&  &  & & -- & &1&[0.2,0.5] & 4 $-$ 13 & 6 $-$ 10  \\
   %&&&&--&&1&[0.25,0.4]& 0.01 $-$ 0.02 & 0.3 $-$ 0.5\\
   &&&&--&0.5$-$1.2&0.1& $>0.7$ \\
\hline
NGC 7793 P13 & [7,8,9,10] & 0.406$-$0.420& 11$-$15  & -- & 5$-$8& 1& $>0.3$ \\
   %&&&&--&&1&[0.10,0.5]& $>10^{3}$ & $>10$ \\
  %& & & & -- &  & 1&[0.25,0.3]& 1 $-$ 16 &  0.7 $-$ 3  \\
    & &&&--&0.5$-$0.8&0.1& $>0.1$ \\

\hline
\end{tabular}
\begin{tablenotes} \item[]\textbf{Notes.} Reference values for $P_\mathrm{spin}$, $\nu_\mathrm{QPO}$ and (in the case of M82\,X-2) $\dot{\mathrm{M}}_\star$, the overall mass-transfer rate from the companion, taken from: [1] \cite{Feng2010}; [2] \cite{Bachetti2022}; [3] \cite{Liu2024}; [4] \cite{RodriguezCastillo2020}; [5] \cite{Imbrogno2024}; [6] \cite{Earnshaw2016}; [7] \cite{Furst2016}; [8] \cite{Israel2017}; [9] \cite{Furst2021}; [10] Imbrogno et al. (in prep). The reported values of the mass-accretion rate $\dot{\mathrm{M}}_{\mathrm{acc}}$ for b=1 have been inferred from the X-ray luminosities (typically in the 0.3--10\,keV band) of the datasets where QPOs have been detected.
%\item[a] Description of Data 3.
%\item[b] Description of Data 6.
\end{tablenotes}
\end{threeparttable}
\end{center}
\end{table*}

\section{Results}
\label{sec:Results}
We apply the model described in Section \ref{sec:Model} and explore the parameter space as detailed in Table\,\ref{tab:model_par} to three PULXs, namely M82\,X-2, M51\,ULX-7 and NGC\,7793 P13, for which both the spin and QPO periods have been detected. 

The parameter space in Table\,\ref{tab:model_par} presents a wide array of combinations leading to a family of solutions with the same QPO and spin. To explore this non-linear parameter space and recover families of solutions that can reproduce the observed spin and QPO frequencies, we first generate $10^{8}$ model realisations using parameter combinations randomly drawn from a flat distribution in log space (apart from $\theta$ as the range spans fewer orders of magnitude than the other parameters) within the ranges given in Table\,\ref{tab:model_par}. We then retain those parameter combinations 
for which both the model-predicted QPO and spin are within the 1$\sigma$ errors of their measured values as reported in Table\,\ref{tab:2}. In the case of M82~X--2, where several QPO measurements exist, we take the average of the individual QPO measurements from Table 1 of \citet{Feng2010}. In the case of M51 ULX-7, the QPO was fitted with two separate Lorentzians in \citet{Imbrogno2024}. Here we assume that the Lorentzian at lower frequency with larger $Q = \frac{\nu}{\Delta\nu} \geq 2.1$ is the QPO, whilst the higher frequency Lorentzian corresponds to the ``broad component''. Finally, in the case of NGC 7793 P13 we found that one Lorentzian was enough to model the QPO (Imbrogno et al., in prep.) and we took the average value of the QPO frequency.

\begin{figure*}
	\includegraphics[width=1\textwidth]{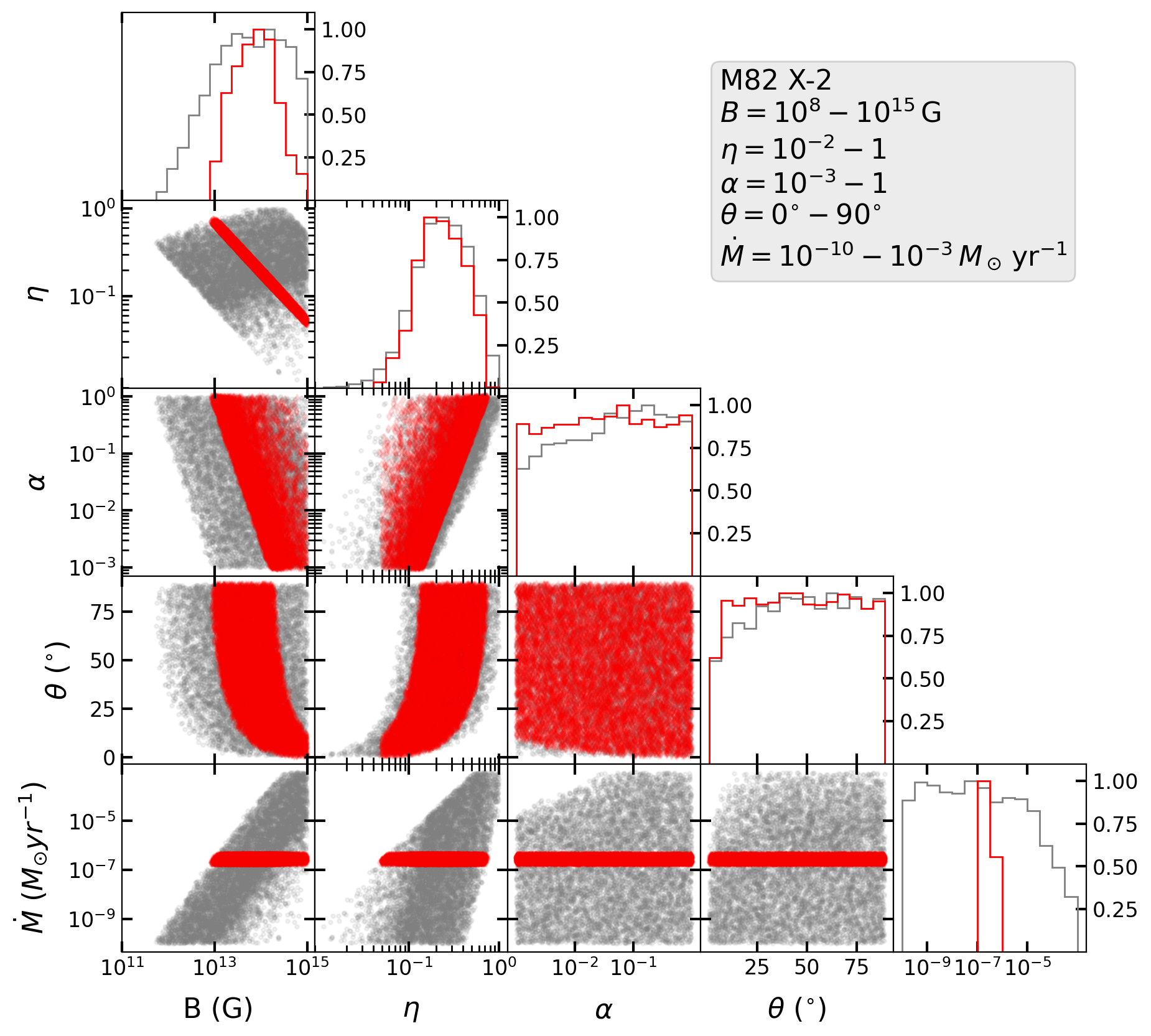}
    \caption{Parameter space of the MDP model with all parameter combinations (in grey circles) producing QPO frequency and spin within the observed errors as noted in Table \ref{tab:2}. All the solutions which also reproduce accretion rate within the range given by the observed $\dot{P}$ are given in red diamonds. Distributions of all parameters are provided, with all being scaled to unity. The explored parameter ranges are noted as also shown in Table \ref{tab:model_par}.
    }
    \label{fig:QPO_model}
\end{figure*}

The model realisations (under the assumption of no beaming i.e. $b = 1$) which reproduced the QPO frequencies and spin periods within their errors are shown in Figure \ref{fig:QPO_model} for M82\,X-2 and in Figure \ref{fig:QPO_model_M51} and \ref{fig:QPO_model_P13} for M51\,ULX-7 and NGC\,7793 P13 respectively, as grey points. Figures \ref{fig:QPO_model}, \ref{fig:QPO_model_M51} and \ref{fig:QPO_model_P13} show a corner plot displaying a parameter sweep through all combinations. 

We first consider the case of M82\,X-2, since it is also the only PULX for which an estimated measurement mass-transfer rate from the donor star ($\dot{M}_{\star}$) has been inferred from the derivative of the orbital period \citep{Bachetti2022} (although note \citet{King2021} associate this signal to stochastic variability). Assuming the orbital period derivative is indeed driven by the mass-transfer rate from the donor, we can reduce the number of family of solutions that can reproduce the QPO and spin of M82\,X-2. The red diamonds in Figure \ref{fig:QPO_model} represent another set of model realisations in which all parameters are randomly drawn from flat distributions as for the grey points, but are additionally constrained by a uniform distribution within the range of $\dot{M}_{\star}$ in Table \ref{tab:2}. With this additional constrain a correlation between $B$ and $\eta$ is observed, where lower $B$ requires higher $\eta$ (Figure \ref{fig:QPO_model}). A similar relation is found between $B$ and $\dot{M}$, where higher $\dot{M}$ provides reasonable constraints for higher $B$, when not considering the $\dot{M}$ constraint from Table \ref{tab:2}. Therefore, limiting $\dot{M}$ for M82\,X-2 yields a smaller set of family of solutions with a lower limit on $B\gtrsim1\times10^{13}$G. Furthermore the observed X-ray luminosity can be explained with no beaming with the assumed mass accretion rate. Nevertheless, we show in Table \ref{tab:2} the effects on the recovered $B$ field assuming a moderate beaming of $b=0.1$. We also explored the effects of moderate beaming, likely due to the geometry of the inner part of the accretion disk and the presence of a wind component often observed in ULXs \citep{Middleton2015, Pinto2016,Pinto2023AN}. It is worth noting that the MDP model does not directly include beaming but only mass accretion rate. Thus in producing the red points in Figure \ref{fig:QPO_model}, $b$ is set to the minimum luminosity required to justify the observed $\dot{P}$. The effect of varying the beaming factor (for $b=0.1$) is shown by the blue diamonds in Figures \ref{fig:QPO_model_M51} and \ref{fig:QPO_model_P13}. This has the effect of lowering the required accretion rate (see Eq. \ref{eq:mdot}), which allows for a factor $\approx3$ lower $B$ fields. The other parameters remain largely unaffected. In either case, the dipolar magnetic field $B$ is found to be larger than 10$^{12}$ G for all three sources.

\section{Discussion and conclusions}
\label{sec:Discussion}

The magnetic torque exerted by the NS on the accretion disc generates a precession of the innermost regions, potentially giving rise to the observed mHz QPOs. Other torques have been invoked to explain these QPOs, as well as the superorbital periodicities observed in several ULXs. For instance, the Lense-Thirring precession model has been applied to the non-pulsating NGC 5408 X-1, to interpret both its QPOs and time lags \citep{Middleton2019}. Other possible mechanisms that can induce precession are asymmetries in the NS with respect to its spin axis \citep[free-body precession, ][]{JonesAnderson2001}, magnetic or self-induced warping \citep{Lai2003,Pfeiffer2004,Pringle1996}, tidal torque from the donor star \citep{Frank2002apa..book.....F} and warping of an inclined disc due to the relativistic Bardeen-Petterson effect \citep{Bardeen1975}. Two or more of these mechanisms could in principle be competing. This would explain, for instance, the presence of (magnetically-induced) QPOs and superorbital periods observed in all three PULXs discussed in this work. \citet{Middleton2018MNRAS.475..154M} discussed the relative strengths of magnetic and Lense-Thirring torques in ULXs, concluding that the magnetic torque would be negligible for $B\sim10^{10-11}$ G and $\alpha \sim 0.01$. However, for larger $B$, as considered here (Table \ref{tab:2}) and larger $\alpha$, the magnetic torque is expected to be more dominant and becomes a viable option for generating QPOs. Another relativistic effect that can be present concurrently with magnetic warping is the Bardeen-Patterson effect \citep{Bardeen1975,King2005MNRAS.363...49K,Ingram2019}. In such a case, an inner tilted disc is warped towards the orbital plane of the system at similar scales as the Lense-Thirring precession. However, a lower limit on the QPO frequency caused by this effect can be set at $\sim3$\,Hz \citep{Fragile2001ApJ...553..955F}. This is significantly higher than the frequencies of the QPOs reported in this work in Table \ref{tab:2}, making it unlikely that the Bardeen-Petterson effect and the magnetically induced warp can counteract one another on the same scale size in the disc.

Despite the complexity of the phenomena that can generate QPOs, in this work we focussed on the effects of a misaligned NS magnetic field to the accretion disc (MDP model) and recovered reasonable family of solutions. In the following Sections, we discuss and outline the most important aspects and limitations of the model when applied to the QPOs in PULXs (Section \ref{subsec:ModelLimitation}), the effects of beaming on the accretion rate of the systems, the implications on the preferred model parameters (Section \ref{subsec:beaming}), the implications of the model on the properties of the PULXs, as well the potential avenues for identifying future candidates of mHz QPOs that can test the MDP model further (Section \ref{subsec:implications}).

\subsection{Limitations of the model}
\label{subsec:ModelLimitation}

Given the number of free parameters in Table~\ref{tab:model_par} any conclusions drawn from the model application should be treated with caution. Apart from the lack of direct observational evidence for the viability of the inferred parameters, it is also important to discuss other assumptions used by the model. One of them is the systems are in spin-equilibrium, such that $r = r_{\mathrm{in}} = r_{\mathrm{M}} = r_{\mathrm{co}}$. This is not necessarily the case for any system with an observed secular spin-up or spin-down \citep{Furst2021}. By analogy however, accreting WDs that are thought to be in spin-equilibrium do display both spin-up and/or spin-down around the equilibrium point \citep{Patterson2020ApJ...897...70P}. Nonetheless, the adopted spin-equilibrium assumption adopted here conveniently removes several parameter degeneracies. As already mentioned, this assumption is likely correct only by M82 X-2 at present - but crucially not when the QPOs were detected. 

We further note that the model is very sensitive to parameters $\alpha$ and $\theta$, whose combinations are degenerate in all ULXs as seen in Figures \ref{fig:QPO_model}, \ref{fig:QPO_model_M51} and \ref{fig:QPO_model_P13}. As shown in Figure \ref{fig:QPO_model} for M82 X-2, high accretion rates rule out low viscosity $\alpha$ and low angles $\theta$.

The MDP model is also very sensitive to changes in $\eta$. However constraining the mass accretion rate significantly constrains $\eta$. The standard value of $\eta \sim 0.5$ used in literature for magnetic accretors \citep{Wang1987,Ghosh1979a,Campana2018} is used as the fiducial value and reproduces reasonable results, for $\alpha \gtrsim 0.1$ and $B$ on the lower end of the range given in Table \ref{tab:2} for most systems. However M82 X-2 has been reported to be out of spin-equilibrium during some of the QPO measurements \citep{Liu2024}. It is thus not surprising that families of solutions with $\eta < 0.5$ can reproduce the observed accretion rate in Table \ref{tab:2}. 

Similarly, NGC 7793 P13 is known to be out of spin-equilibrium (see the long-term evolution of the spin period discussed by \citealp{Furst2021} and \citealp{Furst2025}). Following the example of M82 X-2, $\alpha \sim 0.1 - 0.2$ would require setting $\eta \sim 0.3$ to reproduce the observed accretion rates $\dot{\mathrm{M}}_{\mathrm{acc}}/b$, well below the observational limit for the mass loss of the known donor (see below). Naturally, a higher value of $\alpha$ would then also allow for $\eta = 0.5$.

The present model does not take into account all key aspects of super-Eddington accretion physics \cite[e.g.][]{Shakura1973,Poutanen2007}, but it is interesting nonetheless that it can recover reasonable families of solutions. 
For a NS accreting at super-Eddington rates, the magnetospheric radius, $r_\mathrm{M}$, is usually smaller than $r_\mathrm{sph}$, depending on the magnetic field strength and accretion rate, as also shown in numerical simulations \citep{Chashkina2017,Chashkina2019}, also inferred from the application of the MDP model to the three considered PULXs. Using the equation for spherisation radius from \citet{Poutanen2007} and the measured spin periods of the 3 PULXs considered here, the spherisation radii for the sources considered here are $\sim$ 990 $R_\mathrm{g}$, $\sim$ 165 $R_\mathrm{g}$ and $\sim$ 330 $R_\mathrm{g}$ for M82 X-2, M51 ULX-7 and NGC 7793 P13 respectively, with $R_\mathrm{g}$ being the gravitational radius. However the associated magnetospheric radii based on the observed spins and $r_\mathrm{M} = r_\mathrm{CO}$ are $\sim$ 990 $R_\mathrm{g}$, $\sim$ 1700 $R_\mathrm{g}$ and $\sim$ 450 $R_\mathrm{g}$ for M82 X-2, M51 ULX-7 and NGC 7793 P13 respectively. This suggests that $r_\mathrm{M} \sim r_\mathrm{sph}$ only for M82 X-2. For M51 ULX-7 and NGC 7793 P13 $r_\mathrm{M} > r_\mathrm{sph}$, as also suggested by \citet{Gurpide2021} for high magnetic fields. This would imply the model applied here is self-consistent, at least for M51~ULX7 and NGC~7793~P13 and justifies our assumption of ignoring mass-loss.

\subsection{Beaming factor}
\label{subsec:beaming}

The MDP model requires a mass accretion rate, which can be converted to an observed luminosity assuming a beaming factor (Eq. \ref{eq:mdot}). As already discussed in Section~\ref{sec:Results}, beaming is likely present and should be considered. Whether this beaming is moderate or extreme is still a matter of debate (see e.g. \citealp{Israel2017a} and \citealp{King2020} for both scenarios). 

For M82\,X-2, strong beaming is not required to reconcile the observed X-ray luminosity with the mass accretion rate inferred by \cite{Bachetti2022} through the secular orbital period derivative $\dot{P}_\mathrm{orb}$. In the case of M51 ULX-7 and NGC 7793 P13, the available X-ray data do not allow a $\dot{P}_\mathrm{orb}$ measurement, and therefore an estimate of $\dot{\mathrm{M}}_\star$. However, the latter can be inferred from the observed luminosity, as $L_\mathrm{obs}=L_\mathrm{acc}/b$. Some recent observational results suggest that if the beaming is present, it should be relatively small \citep[see e.g.][]{Israel2017a,RodriguezCastillo2020}.  This is also the case for another PULX, NGC 5907 ULX-1, for which a luminosity of 10$^{41}$ erg s$^{-1}$ has been inferred from its surrounding nebula, suggesting that it is a genuine super-Eddington accretor with no need of (strong) beaming \citep{Belfiore2020}. Additionally, the morphology and strength of He \textsc{ii}, H$\beta$ and [O \textsc{i}] emission regions in the nebula surrounding Ho II X-1 and NGC 1313 X-1 strongly argue against the presence of extreme beaming \citep{Kaaret2004,Gurpide2024a}. Moreover, the inferred mass-transfer from the companion $\dot{M}_{\mathrm{orb}}$ in M82 X-2 shows that the mass available for accretion onto the NS is $\sim$150\,M$_{\odot}$yr$^{-1}$, high enough to account for the observed luminosity (100 times the Eddington limit) without invoking (strong) beaming \citep{Bachetti2022}. Finally, as already emphasised, strong beaming would imply a too small torque on the NS to account for the observed $\dot{P}$. Because of this we explored moderate beaming factors ($>$1/10), resulting in a reduction of the predicted $B$ field by about a factor of $\sim4$. For M82 X-2 and M51 ULX-7 this still poses a lower limit on the $B$ $\gtrsim10^{13}$~G within the MDP model assumptions and $B$ $\gtrsim10^{12}$~G for NGC 7793 P13.

\subsection{Implications of the model and future perspectives}
\label{subsec:implications}

As discussed in Section~\ref{subsec:ModelLimitation}, the MDP model is based on the standard accretion theory for an optically thick, geometrically thin disc \citep{Shakura1973}, with a modification for  the disc thickness to mimic a supercritical accretion disc, %Hence, all the physical properties of the systems accounted for in the model (i.e. viscosity, height, surface density) are for a thin-disc case. 
but without taking into account any further element specific to super-Eddington accretion (winds, thick disc, radiation pressure, etc.). When applied to the sample of PULXs, the model yields reasonable families of solutions with mass-accretion rates, spin and QPO frequencies consistent with those obtained from observations. Within the recovered families of solutions, the inferred dipolar magnetic fields are found to be between a few 10$^{12}$ G and a few 10$^{13}$ G, also consistent with the literature estimates \citep{Furst2016,Israel2017,RodriguezCastillo2020,Bachetti2022}, and of the same order of magnitude to the known X-ray pulsars in HMXBs in our Galaxy and in the Magellanic Clouds. 

A potential implication of the MDP model may be the transitional behaviour of QPOs and spin pulses in M51 ULX-7, where the QPO is observed when the spin signal is absent and vice versa, without significant spectral variations in the source (see \citealp{Imbrogno2024} for details). Within the MDP model, this may suggest that the tilted, precessing inner flow could partially or fully obscure the neutron star’s accretion column. In this case, large values of $\theta$ would cause a drop in pulse coherence, while when $\theta \sim 0$, the alignment minimises obscuration, leading to increased pulse coherence and the absence of QPOs, due to the inner disc symmetry. It is worth clarifying that while the lack of spin pulses is attributed to the precession of the inner disc, $\theta$ relates to the tilt of the outer parts of the disc \citep{Veresvarska2024}. This suggests that the outer disc may also precess to some extent, potentially influencing the overall dynamics of the system, as noted in \citet{Imbrogno2024}. A more detailed investigation of how the outer and inner disc regions interact and contribute to observed variability could provide further insights into the coupling between QPOs, spin pulses, and disc precession.

Further implication arises from the expected observational correlations depending on the assumed disc half-height prescription. As mentioned above, assuming a thin disc from \citet{Shakura1973} yields a positive correlation between the QPO frequency and the accretion rate. On the other hand, the disc half-height prescription more appropriate for a geometrically thick ULX inner disc \citep{Lipunova1999AstL...25..508L} yields a negative correlation. Therefore, with sufficient data on a given QPO in different states, the QPO frequency is expected to change based on the underlying accretion rate, with the variations being driven by the type of inner accretion flow. A QPO with a non-varying frequency throughout different states would thus rule out the MDP model. However, such data is not available, with M51 ULX-7 showing very little variation in QPO frequency over $\sim$ 10 year period where state changes have not been observed \citep{Imbrogno2024}.

In principle, the MDP model could provide a direct way to estimate the magnetic field strength from the QPO and spin frequencies. However, the dependence on other unconstrained model parameters ($\alpha$, $\theta$ and accretion rate) make this challenging. Despite this, new and independent measurements of the magnetic field strength, could lead to a new method to constrain $\eta$ and potentially $\alpha$. 

%% CHECK
Furthermore, with M82 X-2 in spin equilibrium and based on the assumptions that spin equilibrium implies $\eta$ $\sim$ 0.5 and $r_{\mathrm{M}} \sim r_{\mathrm{CO}}$, the MDP model predicts that the QPO should move to lower frequencies for the same spin and accretion rate ($\sim5\times10^{-4}$ Hz assuming the viscosity remains unchanged).

Of all the analysed PULXs, NGC 7793 P13 is the only one with a known companion star \citep[B9Ia class, ][]{Motch2014}. The estimated mass-loss rate of the star can be as high as $\sim10^{-5}$~M$_\odot$\,yr$^{-1}$ \citep{ElMellah2019}. Adopting this value as an upper limit on the mass-transfer rate in ULXs, we can place limits on the expected spin period and magnetic field strength of other candidate PULXs showing mHz QPOs. As an example, 2CXO\,J140314.3+541816 \citep{Urquhart2022} and 4XMM\,J111816.0-324910 \citep{Motta2020} have been suggested to host accreting NSs due to a hard spectrum and large variation in the luminosity, respectively. Using the same model set-up and adopting the fiducial parameters from Table \ref{tab:model_par}, we explore the family of solutions adopting the strict upper limit for the mass-transfer rate of $\dot{M}<$10$^{-5}$ $\mathrm{M}_{\odot}\,\mathrm{yr}^{-1}$. For the QPO in 2CXO\,J140314.3+541816 %\citep{Urquhart2022} 
at 1.35$-$1.92\,mHz we infer $0.5~\mathrm{ms} \lesssim P_\mathrm{spin}\lesssim 8~\mathrm{s}$ and $B > 3 \times 10^{9}$\,G, consistent with the previously suggested $B \sim 10^{10}$G and $P_\mathrm{spin} \sim 5$ms \citep{Urquhart2022}. Similarly for 4XMM\,J111816.0-324910 %\citep{Motta2020} 
we infer $0.5~\mathrm{ms} \lesssim P_\mathrm{spin}\lesssim 4~\mathrm{s}$ and $B \sim 10^{10}$G for QPO frequency $\sim$0.3~mHz. We note that these limits only pertain to the specific case of the fiducial parameters from Table \ref{tab:model_par}. Nevertheless, assuming the magnetically driven precession model as the mHz QPO driver suggests a spin period range for potential new pulsation detections. Obtaining new independent measurements of the dipolar magnetic field of the NSs could help remove some of the degeneracies in the model. This could then potentially open up an avenue of independent constraint on the elusive nature of viscosity $\alpha$.\\

\section*{Acknowledgements}

MI is supported by the AASS Ph.D. joint research programme between the University of Rome "Sapienza" and the University of Rome "Tor Vergata", with the collaboration of the National Institute of Astrophysics (INAF). MV acknowledges the support of the Science and Technology Facilities Council (STFC) studentship ST/W507428/1. SS is supported by STFC grant ST/T000244/1 and ST/X001075/1. 
RA and GLI acknowledge financial support from INAF through grant ``INAF-Astronomy Fellowships in Italy 2022 - (GOG)''. GLI also acknowledge support from PRIN MUR SEAWIND (2022Y2T94C) funded by NextGenerationEU and INAF Grant BLOSSOM. DdM acknowledges support from INAF through "Astrofisica Fondamentale 2022 - Large grant N.16

%%%%%%%%%%%%%%%%%%%%%%%%%%%%%%%%%%%%%%%%%%%%%%%%%%
\section*{Data Availability}
The data analysed in this article are public and can be downloaded from the \xmm\ Science Archive XSA (\texttt{http://nxsa.esac.esa.int/nxsa-web/\#search}) and the High Energy Astrophysics Science Archive Research Center (HEASARC) archive (\texttt{https://heasarc.gsfc.nasa.gov/cgi\-bin/W3Browse/w3browse.pl}).

%%%%%%%%%%%%%%%%%%%% REFERENCES %%%%%%%%%%%%%%%%%%

% The best way to enter references is to use BibTeX:

\bibliographystyle{mnras}
\bibliography{bibliography} % if your bibtex file is called example.bib

% Alternatively you could enter them by hand, like this:
% This method is tedious and prone to error if you have lots of references
%\begin{thebibliography}{99}
%\bibitem[\protect\citeauthoryear{Author}{2012}]{Author2012}
%Author A.~N., 2013, Journal of Improbable Astronomy, 1, 1
%\bibitem[\protect\citeauthoryear{Others}{2013}]{Others2013}
%Others S., 2012, Journal of Interesting Stuff, 17, 198
%\end{thebibliography}

%%%%%%%%%%%%%%%%%%%%%%%%%%%%%%%%%%%%%%%%%%%%%%%%%%

%%%%%%%%%%%%%%%%% APPENDICES %%%%%%%%%%%%%%%%%%%%%

\appendix

\section{Magnetic precession model parameter space for M51 ULX-7 and NGC 7793 P13}

%If you want to present additional material which would interrupt the flow of the main paper,
%it can be placed in an Appendix which appears after the list of references.

\begin{figure*}
	\includegraphics[width=1\textwidth]{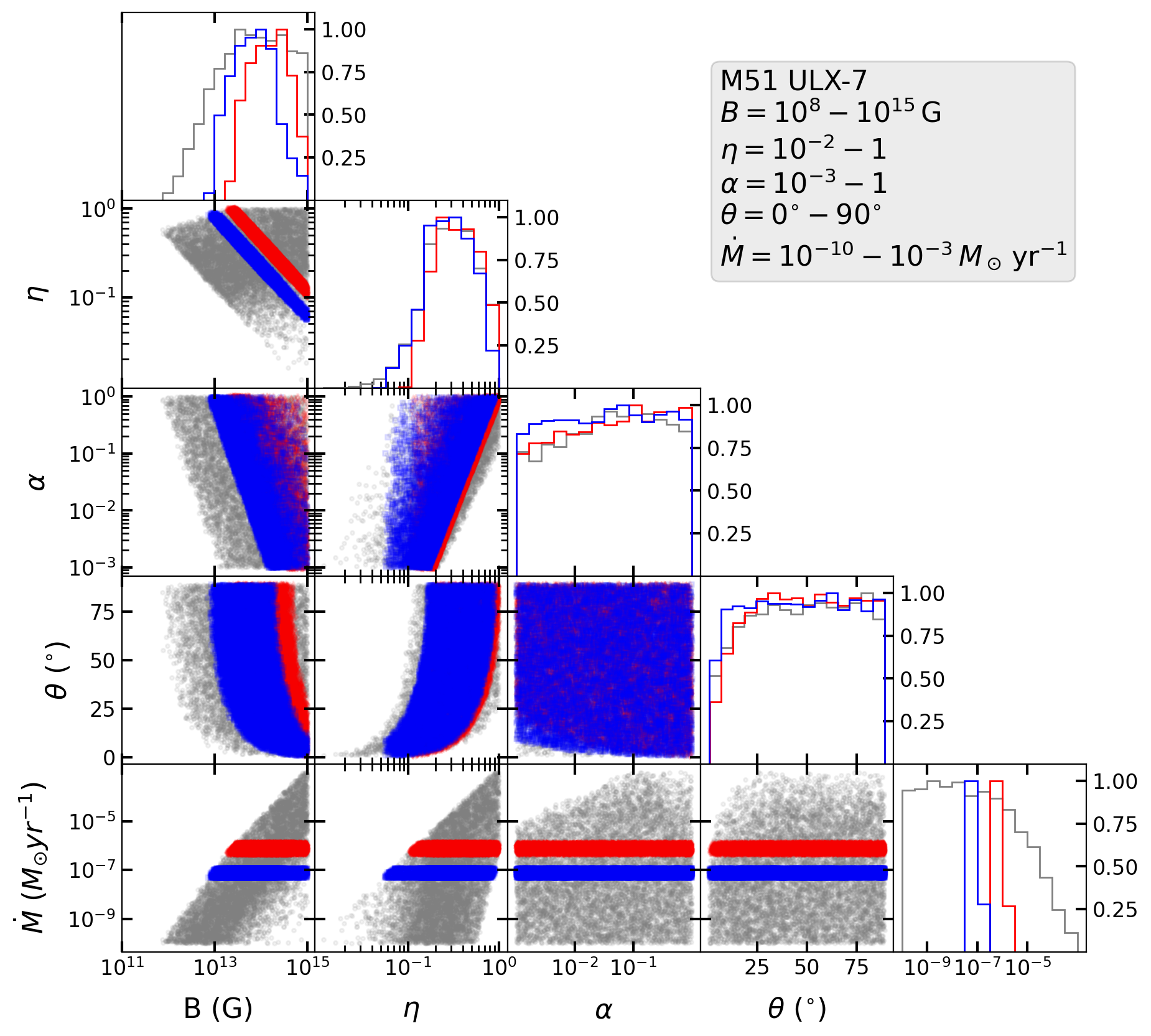}
    \caption{Parameter space of the MDP model for M51 ULX-7 with all parameter combinations (in grey circles) producing QPO frequency and spin within the observed errors as noted in Table \ref{tab:2}. All the solutions which also reproduce accretion rate within the range in Table \ref{tab:2} are given in red diamonds for no beaming (b = 1) and in blue squares for moderate beaming of b = 0.1. Distributions of all parameters are provided, with all being scaled to unity. The explored parameter ranges are noted as also shown in Table \ref{tab:model_par}.
    }
    \label{fig:QPO_model_M51}
\end{figure*}

\begin{figure*}
	\includegraphics[width=1\textwidth]{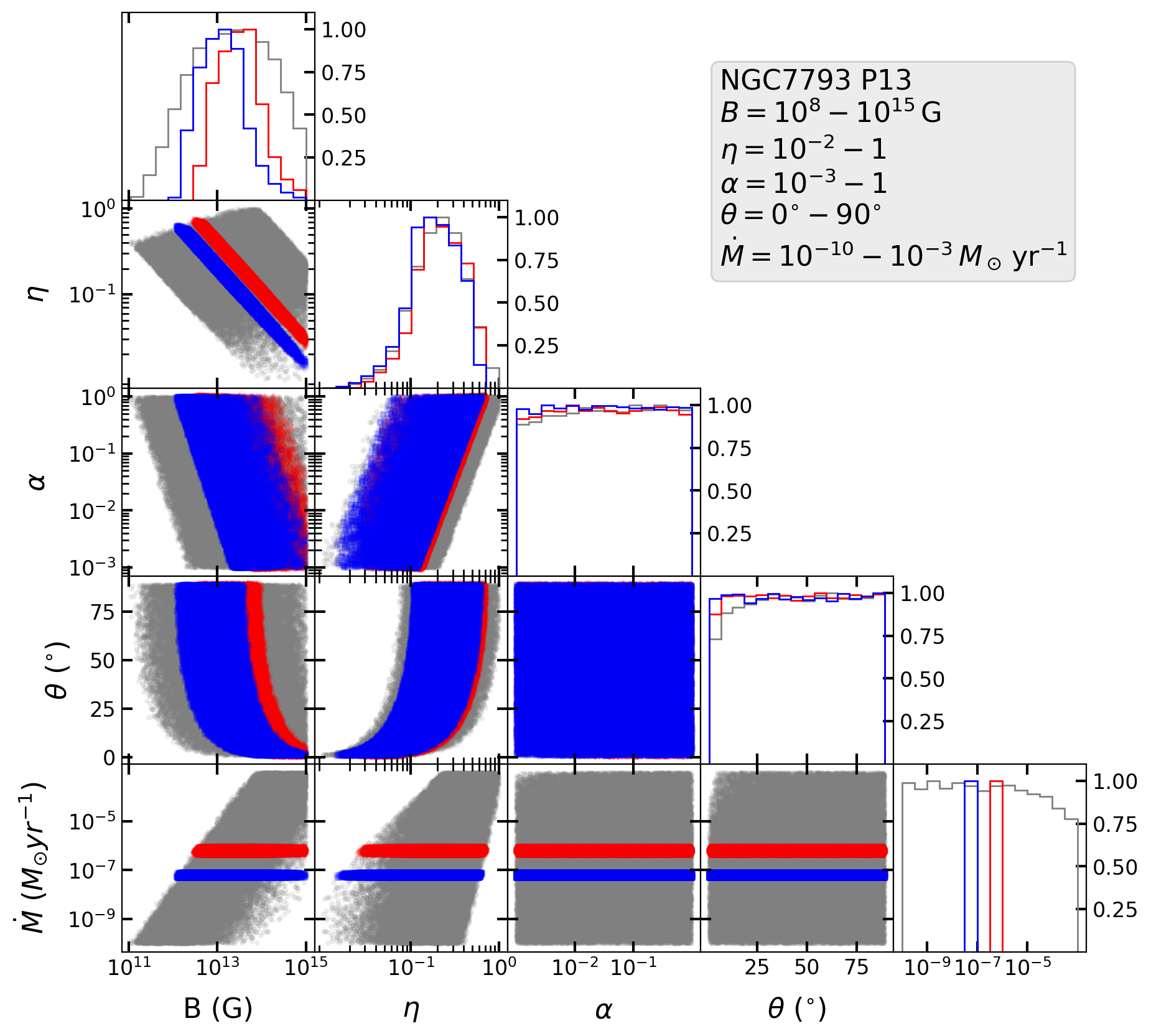}
    \caption{Parameter space of the MDP model for NGC 7793 P13 with all parameter combinations (in grey circles) producing QPO frequency and spin within the observed errors as noted in Table \ref{tab:2}. All the solutions which also reproduce accretion rate within the range in Table \ref{tab:2} are given in red diamonds for no beaming (b = 1) and in blue squares for moderate beaming of b = 0.1. Distributions of all parameters are provided, with all being scaled to unity. The explored parameter ranges are noted as also shown in Table \ref{tab:model_par}.
    }
    \label{fig:QPO_model_P13}
\end{figure*}

%%%%%%%%%%%%%%%%%%%%%%%%%%%%%%%%%%%%%%%%%%%%%%%%%%

% Don't change these lines
\bsp	% typesetting comment
\label{lastpage}
\end{document}